\documentclass[prl,twocolumn,showpacs,superscriptaddress,preprintnumbers,amssymb]{revtex4-2}
\usepackage{newtxtext,newtxmath}
\usepackage{CJK}
\usepackage{fontsize}
\changefontsize{10.5}
\newcommand{\makeCJKtitle}{
    \begin{CJK*}{UTF8}{bkai}
        \maketitle
    \end{CJK*}}
\usepackage{graphicx}
\usepackage{float}
\usepackage[dvipsnames]{xcolor}
\usepackage{dcolumn}
\usepackage{bm}
\usepackage{physics}
\usepackage{comment}
\usepackage{svg}
\usepackage{placeins}
\usepackage[colorlinks,allcolors=Blue,linktocpage]{hyperref}
\usepackage{soul}
\usepackage{multirow}
\DeclareMathAlphabet{\mathbbold}{U}{bbold}{m}{n}

\newcommand{\beq}{\begin{equation}}
\newcommand{\eeq}{\end{equation}}
\newcommand{\beqn}{\begin{eqnarray}}
\newcommand{\eeqn}{\end{eqnarray}}

\newcommand{\p}{\partial}

\newcommand{\SO}{\mathrm{SO}}
\renewcommand{\O}{\mathrm{O}}

\newcommand{\rN}{\mathrm{N}}
\newcommand{\rS}{\mathrm{S}}

\newcommand{\sfigref}[2]{Fig.~\hyperref[#1]{\ref{#1}#2}}

\begin{document}

\title{Fortuitous Universality of Bose-Kondo Impurities}

\author{Abhijat Sarma}

\affiliation{Department of Physics, University of California,
Santa Barbara, CA 93106, USA}
\affiliation{Perimeter Institute for Theoretical Physics, Waterloo, Ontario, N2L 2Y5, Canada}

\author{Zheng Zhou (周正)}

\affiliation{Perimeter Institute for Theoretical Physics, Waterloo, Ontario, N2L 2Y5, Canada}

\affiliation{Department of Physics and Astronomy, University of Waterloo, Waterloo, Ontario N2L 3G1, Canada}

\author{Ryan A. Lanzetta}

\affiliation{Perimeter Institute for Theoretical Physics, Waterloo, Ontario, N2L 2Y5, Canada}



\author{Yin-Chen He}

\affiliation{Perimeter Institute for Theoretical Physics, Waterloo, Ontario, N2L 2Y5, Canada}
\affiliation{C.~N.~Yang Institute for Theoretical Physics, Stony Brook University, Stony Brook, NY 11794-3840}

\begin{abstract}
    \changefontsize{10}\normalsize
    We use the fuzzy-sphere approach to study the Bose-Kondo impurity problem, namely a spin-$S$ impurity coupled to the $(2+1)$-dimensional $O(3)$ Wilson-Fisher CFT (Heisenberg universality class). We demonstrate that for $S=1/2,1,3/2$ the impurity flows to a distinct stable interacting conformal defect for each $S$. Using large-scale exact diagonalization and density-matrix renormalization group methods, we observe integer-spaced defect spectrum consistent with defect conformal symmetry and compute several low-lying defect primary operators as well as the RG monotonic $g$-function. Our findings show that despite sharing the same symmetry and anomaly, Bose-Kondo impurities flow to distinct stable infrared conformal fixed points, which we refer to as \emph{fortuitous universality}. We expect this fortuitous universality to persist for all $S$, extending to $S\rightarrow\infty$, with each spin-$S$ impurity flowing to its own stable infrared conformal fixed point.
\end{abstract}

\makeCJKtitle
\emph{Introduction.} A cornerstone of the renormalization-group (RG) paradigm~\cite{Wilson:1974mb} is universality: microscopically distinct systems can exhibit identical long-distance behavior controlled by a small set, sometimes a unique one, of infrared (IR) fixed points, determined primarily by symmetry, anomaly, and dimensionality. A prominent example is the $O(3)$ Wilson-Fisher (WF) universality class~\cite{WilsonFisher} in $(2+1)$ dimensions, which governs the critical behavior of many anomaly-free $SO(3)$-invariant quantum spin models, regardless of whether the microscopic spins are $1/2$, $1$, $3/2$, or higher~\cite{Matsumoto2001}. Universality provides a powerful organizing principle for IR physics, sharply constraining the landscape of IR fixed points within a given symmetry and anomaly class.

Universality also governs the response of a gapless system to a localized impurity or defect, where localized degrees of freedom interacting with a critical bulk can generate RG flows and IR fixed points distinct from those of the bulk. An early example is the Kondo effect \cite{Kondo:1964nea}, which explains how the screening of magnetic \emph{Kondo} impurities in metals produces the experimentally-observed low temperature resistance minimum in dilute metallic alloys \cite{DEHAAS19341115}. The Kondo problem can be solved essentially exactly using methods such as the Bethe Ansatz~\cite{Andrei:1982cr,exact_res_mag_alloys}  and boundary conformal field theory (CFT)~\cite{Affleck:1995ge}, revealing a rich structure of impurity fixed points governed by screening. In particular, multichannel Kondo systems can exhibit distinct IR fixed points for different impurity spins when the impurity is overscreened by multiple channels, with the multiplicity of IR fixed points tied to the number of screening channels~\cite{Affleck:1990iv}.

The Kondo impurity problem admits a natural generalization to spin impurities coupled to bosonic gapless systems, such as quantum or classical magnets at criticality, which we refer to as \emph{Bose-Kondo} impurities~\cite{sachdev1999quantum,Sachdev_QMC,Cuomo_2022,Defect_anomalies,Whitsitt_2017, Chen_2018, Chen_2019,Defect_anomalies}. Here, we study Bose-Kondo impurities in the $O(3)$ WF CFT, one of the most extensively studied and widely realized $(2+1)$-dimensional CFTs. We uncover that impurities with different spin $S$, even when they share the same symmetry and anomaly, flow to distinct stable IR conformal defect fixed points. In contrast to multichannel Kondo systems, this multiplicity arises without introducing additional bulk screening channels, and instead emerges intrinsically from a single strongly interacting critical bath. We refer to this phenomenon as \emph{fortuitous universality} \footnote{We distinguish fortuitous universality from the notion of \emph{conformal manifold}, wherein there is a manifold of CFTs connected by exactly marginal perturbations. In our case there is no perturbation connecting the different dCFTs.}: the existence of multiple, potentially infinite, IR fixed points sharing the same symmetry, anomaly, and stability, yet exhibiting distinct critical behavior characterized by different critical exponents.

Impurities, like bulk systems, often exhibit emergent conformal invariance at RG fixed points, and are consequently described by defect CFT (dCFT), which generalizes the notion of boundary CFT (bCFT) to arbitrary patterns of broken bulk conformal invariance \cite{Cardy:1984bb,Cardy:1991tv,Lewellen:1991tb, McAvity:1995zd,Billo2013,Billo_2016}. dCFT therefore provides a natural and powerful language for describing Bose-Kondo impurity fixed points, generalizing the successful application of bCFT to the standard Kondo problem. The universal properties of Bose-Kondo impurities are encoded in the dCFT through quantities such as the spectrum of defect operators and the $g$-function~\cite{Affleck_gtheorem,Friedan:2003yc,Casini:2016fgb,Cuomo:2021rkm}, which is an RG monotone quantifying the effective number of degrees of freedom of the impurity fixed point relative to the bulk. The Bose-Kondo impurity problems for spin $S=1/2,1$ have been studied previously in lattice models using the Monte Carlo method~\cite{Sachdev_QMC,troyer_2002,hoglund2007anomalouscurieresponseimpurities}. While some universal quantities such as the spin correlation exponent (only for $S=1/2$) and susceptibility have been measured, clear evidence supporting emergent conformal invariance has not yet been established in either case. Additionally, impurities with $S > 1$ have not yet been simulated, and in all cases a detailed understanding of the impurity universality classes is lacking.

Recent advances in studying $(2+1)$-dimensional CFTs using the fuzzy-sphere regularization~\cite{Zhu2022,He2026Mar}, including its application to dCFTs \cite{Hu2023Aug,Zhou2024Jan,Cuomo2024}, now enable direct access to the conformal structure of Bose-Kondo impurities. We couple the fuzzy-sphere $O(3)$ WF model \cite{Dey2025} to impurity spins located at the north and south poles, which allows us to access the dCFT via the state-operator correspondence. Using exact diagonalization (ED) and density-matrix renormalization group (DMRG), we study impurity spins $S=1/2, 1, 3/2$ and find clear evidence of emergent conformal symmetry at their IR fixed points. We extract the defect operator spectrum and the $g$-function, demonstrating that these impurities flow to distinct dCFTs. Together with analytic evidence~\cite{Cuomo_2022} and preliminary ED results for higher impurity spins $S=2, 5/2$, we expect that Bose-Kondo impurities in the $O(3)$ WF CFT flow to a distinct stable dCFT for each half-integer and integer $S$.  The fixed points in this infinite family are sharply distinguishable: increasing $S$ leads to a growing number of independently diverging susceptibilities (i.e., operators with scaling dimensions smaller than $1/2$~\cite{Affleck:1995ge}), providing a concrete experimental signature of this realization of fortuitous universality.

\emph{Bose-Kondo impurities.} There are various ways to model the Bose-Kondo impurities in the $O(3)$ WF CFT. In a lattice model, one can either introduce an impurity spin coupled to the bulk $O(3)$ universality class or realize the impurity via lattice vacancies~\cite{Sachdev_QMC}. In field theory, the spin-$S$ impurity  can be modeled as a $SO(3)$ vector $\vec n(z)$ living in the $0+1$ dimension, where $S$ is specified by the Berry phase action, also known as the Wess-Zumino-Witten (WZW) term, $\mathcal S_{\mathrm{WZW}}[\vec{n}] = iS\int d\tau \int_{0}^1 du \; \vec{n} \cdot \p_u\vec{n} \times \p_z \vec{n}$. We place the impurity at $r=0$ and have it interact with the bulk $O(3)$ vector field $\vec \phi$, giving the action
\begin{align}
\label{eqn:bk_WZW}
    \mathcal S_{\mathrm{bulk}}[\vec\phi] + \mathcal S_{\mathrm{WZW}}[\vec{n}] + J_{\mathrm{imp}}\int dz \; \vec n(z) \cdot \vec\phi(z, r=0).
\end{align}
We note that one can equivalently formulate it in the language of line defect operators~\cite{Cuomo_2022}. From the field theory, it is clear that the bulk $O(3)$ symmetry is explicitly broken down to the $SO(3)$ symmetry by the impurity. Especially, the broken improper $Z_2$ of bulk $O(3)$ flips the sign of impurity coupling term $\vec n \cdot \vec \phi$, so the sign of $J_{\mathrm{imp}}$ will not affect the physics. 

A key question is whether a given defect is screened in the infrared (IR), meaning that the IR fixed point is the transparent defect, i.e.~a clean bulk without any defect. A helpful compass in answering this question is the irreversibility of line defect RG flow, otherwise known as the $g$-theorem, which dictates that there exists an RG monotone known as the $g$-function that decreases along RG flow triggered by a relevant perturbation, i.e.~$g_{\text{UV}} > g_{\text{IR}}$ with $g_{\text{UV}}, g_{\text{IR}}$ the $g$-function of the UV and IR defect fixed points, respectively \cite{Affleck_gtheorem,Friedan:2003yc,Casini:2016fgb,Cuomo:2021rkm}. In particular, this means that if the ultraviolet (UV) fixed point is the transparent defect, which has $g_{\text{UV}} = 1$, a defect created by triggering RG flow by a relevant operator cannot be screened. 

However, in our cases, we have $g_{\mathrm{UV}}=2S+1$, because that is the Hilbert space dimension of the spin-$S$ impurity, which is decoupled from the bulk in the UV. The fate of defects with $g_{\text{UV}} > 1$ is somewhat more subtle, since the $g$-theorem alone does not forbid screening. However, another way to definitively exclude screening is to consider an impurity that carries an 't Hooft anomaly of a bulk symmetry $G$, which for a line defect is associated with the line defect endpoint realizing $G$ projectively. This implies that the spin-$S$ defects cannot be fully screened for half-integer $S$. On the other hand, for integer $S$, screening is neither prohibited by the $g$-theorem nor by 't Hooft anomaly; whether these defects are screened or not is thus beyond the scope of these general arguments.

Another possibility is that the nontrivial fixed points discussed above may not be proper dCFTs in which the conformal symmetry of the bulk is actually broken to a smaller subgroup. In particular, the anomalies can also be satisfied with the bulk criticality remaining intact and a residual decoupled impurity degree of freedom, referred to as ``topological quantum mechanics" (TQM) in \cite{Defect_anomalies}; this is exactly the scenario realized in the UV when the impurity spin is totally decoupled~\footnote{In the standard Kondo problem, this is called underscreening.}. For the $S=1/2$ defects this scenario is forbidden in the IR by the $g$ theorem, but for $S>1/2$ it is an open question as to whether these nontrivial defects are dCFTs or TQM. Further, it is unclear whether the defect fixed points of different $S$ are distinct or not; for example, \textit{a priori} the $S=3/2$ defect could potentially flow to the $S=1/2$ defect.

The Bose-Kondo defects in the large-$S$, small $\epsilon=4-d$ limit can be studied perturbatively, in which they flow to genuine dCFTs that are IR stable. In particular, in \cite{Cuomo_2022} it is predicted that the large-$S$ Bose-Kondo defect flows to the pinning field defect, in which the $O(3)$ symmetry is explicitly broken to $O(2)$, with a free spin-$S$ degree of freedom that restores the full $O(3)$ symmetry. In other words, the large-$S$ impurity acts like a Zeeman field for $\vec{\phi}$ whose direction fluctuates freely. This gives strong predictions for the spectrum and $g$-function of the Bose-Kondo defects for large-$S$, but it is unclear if these conclusions hold for the small $S$ values in $2+1$ spacetime dimensions considered in this work.

\begin{figure}
    \includegraphics[width=0.48\textwidth]{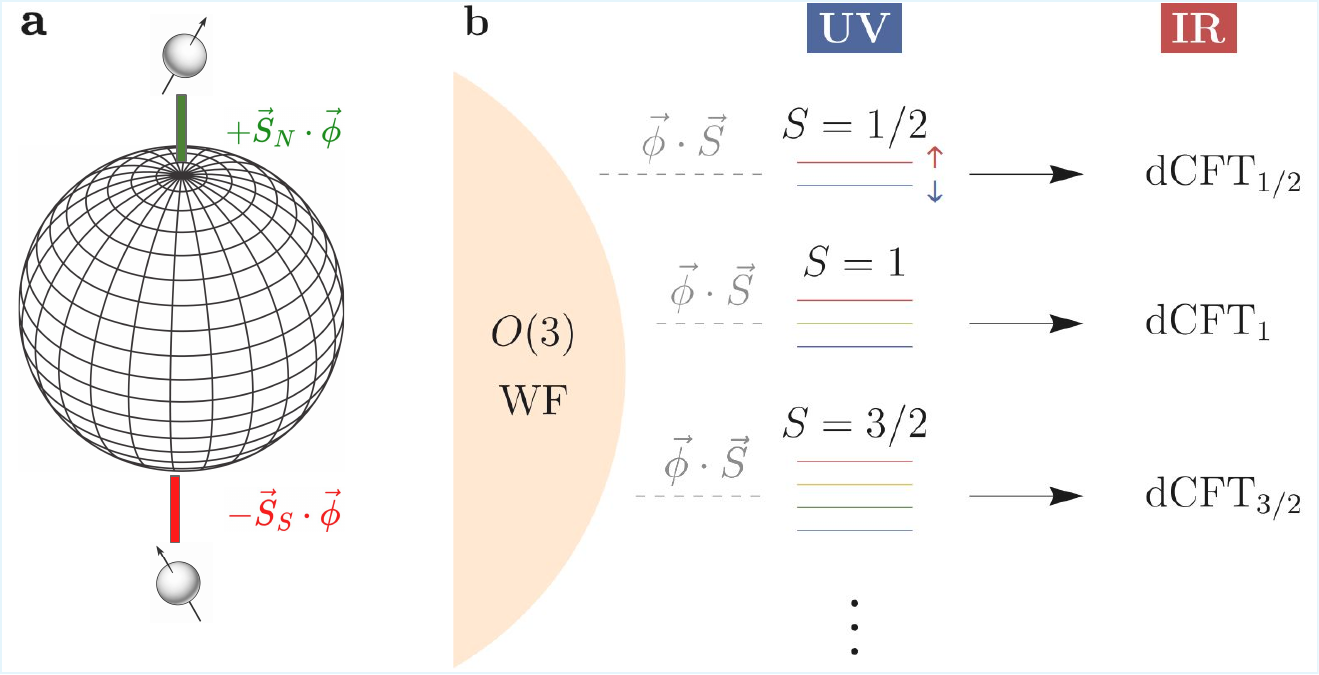}
    \caption{\textbf{a.} \emph{Fuzzy Sphere Setup.} Fermions in the lowest Landau level on $S^2$ interact in a manner that realizes the $O(3)$ WF CFT in the thermodynamic limit. Spin-$S$ impurities are introduced at each pole, coupling to the bulk in a manner that preserves $SO(3) \subset O(3)$ global symmetry. \textbf{b.} \emph{Fortuitous Universality.} Each impurity flows to a distinct stable dCFT in the IR descending from the same bulk criticality, a phenomenon we expect to persist for all $S$. Despite an RG flow from $\text{dCFT}_S$ to $\text{dCFT}_{S'}$ for $S'<S$, $S-S' \in \mathbb{Z}$ being allowed by the $g$-theorem and anomaly, no such flow exists. We refer to such a scenario as fortuitous universality. }
    \label{fig:fortuitous}
\end{figure}

\emph{Fuzzy sphere model.} We now apply the fuzzy sphere regularization~\cite{Zhu2022} to study the Bose-Kondo impurity problem. This approach considers a quantum critical model of interacting fermions on a sphere in the presence of a magnetic monopole projected to the lowest Landau level \cite{Haldane_pseudopot,Ippoliti:2018ojo}. The fuzzy sphere models for $O(3)$ WF CFT have been studied~\cite{Han2023Dec,Dey2025,Guo2025}, and we will utilize the model constructed in Ref.~\cite{Dey2025}, rewritten in the conventions of Ref.~\cite{Guo2025,Fuzzified}. The model has four flavors of fermions $\psi_{\alpha=0,1,2,3}(\bm x)$, where $\psi_{\alpha=1,2,3}(\bm x)$ forms an $O(3)_{\mathrm{int}}$ vector and $\psi_0(\bm x)$ is an $O(3)_{\mathrm{int}}$ singlet. On the fuzzy sphere, i.e. the lowest Landau level, we have 
\begin{equation}
\psi^\dagger_\alpha(\bm x) = \frac{1}{R} \sum_{m=-q}^{q} Y_{qm}^{(q)}(\bm x) c^\dagger_{\alpha,m}
\end{equation} 
where $Y^{(s)}_{lm}$ are the \textit{monopole harmonics} \cite{Monopole_harmonics}. $c^\dagger_{\alpha,m}$ are fermions following canonical anti-commutation relations. We let $N_{m}=2q+1$ denote the number of orbitals per flavor and set $R=\sqrt{N_m}$ as the radius of the fuzzy sphere. We fix the total filling to be a quarter, equivalent to one flavor of fermions being fully filled on average. 

To define the Hamiltonian, we introduce density fields given by bilinears of $\psi_\alpha$, which play the role of the spin operator in the familiar lattice model. Specifically, we define (1) $n_{V,i}(\bm x) = \frac{1}{\sqrt{2}}(\psi_i^\dag(\bm x) \psi_0(\bm x) + \psi_0^\dag(\bm x) \psi_i(\bm x))$ for $i=1,2,3$, which is an $O(3)_{\mathrm{int}}$ vector serving as the order parameter of the WF transition; and (2) $n(\bm x) = \psi^\dag_\alpha(\bm x) \psi_\alpha(\bm x)$ and $n_0(\bm x) = \psi^\dag_0(\bm x) \psi_0(\bm x)$, both of which are $O(3)$ singlets. The bulk Hamiltonian is then given by
\begin{multline}
    H_{\mathrm{bulk}} = \int d^2 {\bm x}_1 d^2 {\bm x}_2 \,\Big[U(|{\bm x}_{12}|) n({\bm x}_1)n({\bm x}_2) \\ - V(|{\bm x}_{12}|) \vec{n}_V({\bm x}_1) \cdot \vec{n}_V({\bm x}_2) - h n_0({\bm x}_1)\delta^2({\bm x}_{12})\Big]
\end{multline}
The interactions $U(r), V(r)$ are chosen to be short ranged, and the details are given in Appendix~\ref{app:num_dets}. Depending on the values of the relative chemical potential $h$, we would have either a $O(3)_{\mathrm{int}}$ disordered phase where all the fermions are in the singlet flavor $\alpha=0$, or a $O(3)_{\mathrm{int}}$ spontaneous breaking phase where the fermions form a coherent state between the $\alpha=0$ and $\alpha=1,2,3$ flavors~\cite{Guo2025}. The $O(3)$ WF transition occurs at a critical $h$, as given in Appendix~\ref{app:num_dets}. The model preserves a $SO(3)_\text{rot}\times O(3)_\text{int}$ symmetry. The spacetime parity, i.e.~the improper $\mathbb{Z}_2$ of $SO(3)_\text{rot}$, is not present.

Finally, we introduce impurity spins to couple to the bulk $O(3)$ WF CFT. To study the state-operator correspondence of the defect operators, the impurity spins are placed at the north and south poles of the sphere~\cite{Hu2023Aug}, interacting with the bulk $O(3)_{\mathrm{int}}$ vector $n_V(\bm x)$ in a $SO(3)_{\mathrm{int}}$ invariant manner, as described by the Hamiltonian,
\begin{equation}
    H = H_{\mathrm{bulk}} + J_{\mathrm{imp}} \vec{n}_V(\rN) \cdot \vec{S}_{\rN} - J_{\mathrm{imp}} \vec{n}_V(\rS) \cdot \vec{S}_{\rS}
    \label{eq:hmt_fs}
\end{equation} 
where $\vec{S}_{\rN,\rS}$ are spin operators acting on the north/south pole spin-$S$ impurity, i.e., a $(2S+1)$-dimensional Hilbert space. We highlight that impurity couplings have opposite signs at the north and south poles. This can be understood by noting that the semi-infinite defects in the northern and southern hemispheres are related by CPT conjugation~\cite{Cordova_2016}, which flips the sign of the defect spin.

The coupling to the impurity breaks the symmetry to $(\SO(2)_\text{rot}\times \SO(3)_\text{int})\rtimes\mathbb{Z}_2$. The presence of the line defect breaks $\SO(3)_\text{rot}$ to $\SO(2)_\text{rot}$, and $\O(3)_\text{int}$ to $\SO(3)_{\text{int}}$. The improper $\mathbb{Z}_2$'s of $\O(2)_\text{rot}$ and $\O(3)_\mathrm{int}$ are reduced to their diagonal $\mathbb{Z}_2$, which acts as the combination of a $\pi$-rotation about the $y$-axis and a sign flip of $\SO(3)_{\text{int}}$ vectors. As a result of these symmetry, each local defect operator must carry a transverse spin $l^z$ under the rotation about the defect line and a $SO(3)_\text{int}$ spin $s$.

\begin{table}[b]
    \centering
    \renewcommand{\tabcolsep}{10pt}
    \begin{tabular}{cc|lll}
        \hline\hline
        & & \multicolumn{3}{c}{$S$} \\
        & $\Delta_{\hat{O}}$ & $1/2$ & $1$ & $3/2$ \\
        \hline
        \multirow{7}{*}{$\hat{O}$} &$\hat{p}_1$ & $0.21(3)$ & $0.118(17)$ & $0.067(12)$ \\
        & $\hat{p}_2$ & $0.72(4)$ & $0.375(17)$ & $0.227(12)$ \\
        & $\hat{p}_3$ &           & $0.93(2)$ & $0.50(2)$ \\
        & $\hat{p}_4$ &           &        & $1.08(2)$ \\
        & $\hat{q}$   & $1.03(5)$ & $1.06(3)$ & $1.04(2)$ \\
        & $\hat{S}$   & $1.82(4)$ & $1.73(2)$ & $1.65(5)$ \\
        \cline{2-5} 
        & $\hat{\phi}_{a0}$ & $0.06(5)$ & $0.09(3)$ & $0.11(3)$ \\
        \hline
        \multicolumn{2}{c|}{$g$-function} & $1.47(5)$ & $2.01(1)$ & $2.59(2)$ \\
        \hline \hline
    \end{tabular}
\caption{\emph{Bose-Kondo dCFT Data.} For each value of $S$ we find a distinct dCFT, showing their $g$-function and some low-lying primaries with $l^z=0$. $\hat p_s$ has $\SO(3)_{\text{int}}$ spin $s$, while $\hat q$ and $\hat S$ have $s=0$. The defect creation operator $\hat{\phi}_{a0}$ carries $l^z=0$ and $s=S$. The data and error bars reported here obtained from a finite-size extrapolation. The dCFT data approaches the analytical large-$S$ predictions, discussed more in the main text.}
\label{tab:data_summary}
\end{table}

\emph{Results.} We use exact diagonalization (ED) and DMRG for numerical calculations for the model~\eqref{eq:hmt_fs} up to $S=3/2$. We are able to reach system sizes up to $N_m = 10$ for $S=1/2$ and $N_m=9$ for $S=1,3/2$ with ED, and $N_m = 15$ with DMRG. The maximum bond dimension is $\chi=5000$ for DMRG. We have also checked $S=2,5/2$ qualitatively up to $N_m=6$ by ED. For a conformal defect, the state-operator correspondence yields that each defect operator $\hat{O}$ corresponds to a fuzzy-sphere eigenstate. The state carries the same transverse spin $l^z$ and the $SO(3)_\text{int}$-spin $s$ as the operator, and the scaling dimension can be extracted from the excited energy 
\begin{equation}
    E_{\hat{O}}-E_0=\frac{v}{R}\Delta_{\hat{O}}
\end{equation}
up to subleading contributions from the irrelevant operators, where $E_0$ is the ground state energy (in the presence of the defect), and the speed $v$ of light is a model-dependent constant to be determined. 

\begin{figure}[t]
    \centering
    \includegraphics[width=0.8\linewidth]{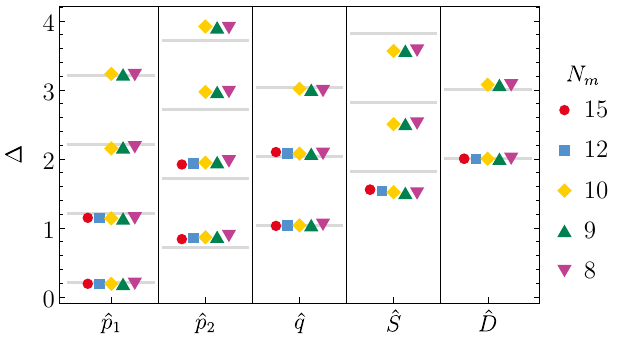}
    \caption{The conformal multiplets of various low-lying primaries in the $S=1/2$ Bose-Kondo defect. The gray bars denote the values anticipated with conformal symmetry. For higher states, only ED data up to $N_m=10$ are available.}
    \label{fig:mult}
\end{figure}

Across a wide range of $J_\text{imp}$ values, we find a clear signature of emergent conformal symmetry. As in any conformal line defect CFT, the spectrum contains a displacement operator $\hat{D}$ with $l^z=1$, $s=0$, and $\Delta_{\hat{D}}=2$ that originates from the broken translation symmetry~\cite{Billo_2016}. We use this protected scaling dimension as the calibrator to obtain $v$. The operators organize into integer-spaced conformal tower --- from a defect primary $\hat{O}$, acting spatial derivatives along the defect line gives its descendants $\partial_0^n\hat{O}$ with the same quantum numbers $l^z$ and $s$ and scaling dimension $\Delta_{\hat{O}}+n$. For all $S=1/2,1$ and $3/2$, we find that all the low-lying states organize into multiplets, and all the descendants of the lowest primaries can be identified (Fig.~\ref{fig:mult}). In each multiplet, the differences between the scaling dimensions are approximately integer and improve as the system size increases. Hence, for all the $S$ being computed, the fuzzy-sphere model flows to a defect CFT. 

Having verified conformal symmetry, we now examine the spectra of the defect primaries summarized in Table~\ref{tab:data_summary}. We find sharp distinctions between the spectra for different $S$, confirming that each $S$ flows to a distinct dCFT. In particular, the spin-$S$ impurity has $2S$ very low-lying primaries ($\Delta \lesssim 0.5$), denoted as $\hat{p}_s$ ($s=1,\dots,2S$), which are $\SO(2)_\text{rot}$ scalars ($l^z=0$) and carry $\SO(3)_{\text{int}}$ spin $s$. This is physically significant because it provides an experimentally accessible way to detect these dCFTs. Specifically, one way to measure critical exponents in a quantum phase transition is through the temperature dependence of the susceptibility $\chi_{\hat O}$ (e.g. spin susceptibility). For the ($0+1$-dimensional) impurity, $\chi_{\hat O} \sim T^{2\Delta_{\hat O}-1}$~\cite{Affleck:1995ge}, and $\chi_{\hat O}$ diverges at low temperature when $\Delta_{\hat O}<0.5$. Therefore, a physical distinction between these dCFTs at different $S$ is the number of diverging susceptibilities, which increases with $S$. Furthermore, the lowest defect singlet $\hat{S}$ with $l^z=0$ and $s=0$ is irrelevant, $\Delta_{\hat{S}}>1$ for all computed $S$, confirming that each dCFT is stable under all symmetry-preserving perturbations. 

Our operator spectrum is qualitatively consistent with perturbative studies of the large-$S$ Bose-Kondo impurity in $4-\epsilon$ dimensions~\cite{Cuomo_2022}. It was found that the $S=\infty$ impurity flows to a pinning-field defect with a decoupled spin-$S$ degree of freedom. Specifically,  $\hat q$ ($l^z=0, s=1$) in our operator spectrum has a scaling dimension close to 1, which will become the ``tilt" operator with a protected scaling dimension $\Delta_{\hat{q},\text{pin}} = 1$ in the $S=\infty$ limit. Further, the scaling dimension of each $\hat{p}_{s}$, $s=1,\dots,2S$ is monotonically decreasing towards 0 as $S$ increases, corresponding to the $(2S+1)^2$ degenerate ground states of a dangling spin.

\begin{figure}[b]
    \centering
    \centering
    \includegraphics[width=0.49\linewidth]{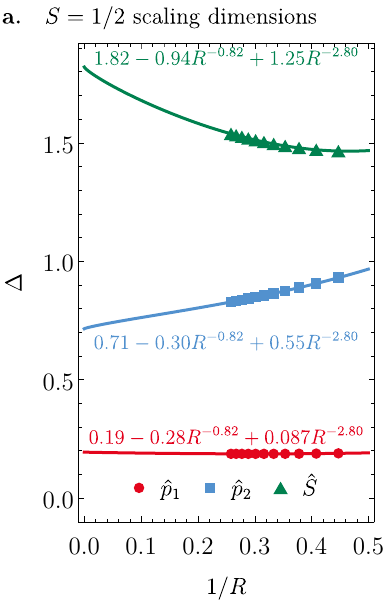}
    \includegraphics[width=0.49\linewidth]{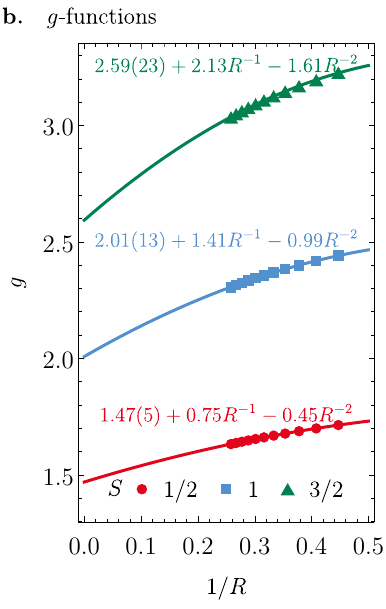}
    \caption{The finite-size extrapolation of various quantities. (a) the scaling dimensions of the defect primaries $\hat{p}_1$, $\hat{p}_2$ and $\hat{S}$ of the $S=1/2$ Bose-Kondo defect; (b) the $g$-functions of the $S=1/2,1$ and $3/2$ Bose-Kondo defect.}
    \label{fig:fss}
\end{figure}

In the calculation, we choose $J_{\mathrm{imp}}$ that optimizes the integer spacing to minimize the finite-size effect. We extrapolate the scaling dimensions through a finite-size scaling with an ansatz taking into account the two lowest lying irrelevant defect primaries (Fig.~\ref{fig:fss}a)~\footnote{We do not include subleading contributions from the bulk as the subleading bulk perturbations are already fine-tuned to zero~\cite{Dey2025}} 
\begin{equation}
    \Delta_{\hat{O}}(R) = \Delta_{\hat{O}} + \frac{c_{\hat{S}}}{R^{\Delta_{\hat{S}}-1}} + \frac{c_{\hat{S}'}}{R^{\Delta_{\hat{S}'}-1}}
    \label{eq:finite_size_scaling}
\end{equation}
where $\hat{S}'$ is the second singlet with scaling dimension around $3$. The results for key operators are listed in Table~\ref{tab:data_summary}. In Appendices \ref{app:dim_fss} and \ref{app:data}, we give the detail for the fitting and show that this ansatz successfully capture the finite-size effect.

Besides the defect operators, we study the defect-creation operators by coupling only the north pole to an external spin-$S$, and extracting the scaling dimension through state-operator correspondence~\cite{Zhou2024Jan}. We find the lowest defect-creation operator carries $SO(3)_\text{int}$ spin-$s$ for $S=1/2,1$ and $3/2$; for half-integer $S$, the creation operator carries a projective representation of $SO(3)_\text{int}$. The scaling dimensions increase with $S$. 

We extract the $g$-function \cite{Affleck_gtheorem} for each $S$ through taking wavefunction overlaps between different defect configurations~\cite{Zhou2024Jan} (See Appendix \ref{app:g_fac}). We extrapolate to the thermodynamic limit through an ansatz $g(R) = g + a R^{-1} + bR^{-2}$ (Table~\ref{tab:data_summary}, Fig.~\ref{fig:fss}b). The results are consistent with the bound $1<g<g_{\mathrm{UV}}=2S+1$ following the $g$-theorem.

\emph{Discussion.} In conclusion, in this work we analyzed Bose-Kondo impurities in the $O(3)$ WF CFT. We showed that for each value of impurity spin $S=1/2,1,3/2$ the impurity flows to a distinct stable conformal defect. In each case we find that the $g$-function satisfies $1<g<2S+1$. We expect that these qualitative aspects of the Bose-Kondo impurities will hold for every value of $S$, giving a seemingly infinite family of nontrivial defect CFTs descending from the $O(3)$ WF bulk criticalities. We extract the scaling dimensions of low-lying operators in each case and show clear evidence of emerging conformal symmetry.

A natural follow-up to our work is to study spin impurities in the $O(2)$ WF CFT, which are referred to as the Halon impurity in the literature \cite{Whitsitt_2017, Chen_2018, Chen_2019,Defect_anomalies}. The Halon impurity arises quite naturally upon doping a critical Bose-Hubbard model at a single site, and features interesting phenomenology including charge fractionalization. This impurity is conjectured to be dual to the \textit{$\pi$-flux vortex line} of the Abelian-Higgs model, which introduces a gauge holonomy of $\pi$ around the defect worldline \cite{Defect_anomalies}. Our preliminary numerical results indicate that the $S=1/2$ Halon impurity also flows to an IR stable dCFT, but it is unclear if higher $S$ would also flow to new stable dCFTs. Another interesting direction is to study conformal defects, such as Wilson lines, in gauge theories, especially given that several such CFTs have been realized on the fuzzy sphere~\cite{Zhou2025Jul,Zhou2023,Zhou2024Oct,Huffman2026}. One particularly intriguing arena is their role in the duality conjecture of Chern-Simons-matter theory~\cite{Zhou2025Jul}, where the electric line (Wilson line) can be dual to the magnetic line ($\pi$-flux vortex line of the gauge field).

Our work also paves the way for future studies of Bose-Kondo impurities in the $O(3)$ WF CFT with numerical conformal bootstrap \cite{Poland:2018epd}, following Ref.~\cite{Lanzetta:2025xfw} and earlier works applying numerical bootstrap to the study of line defects \cite{Gaiotto:2013nva,Liendo:2018ukf,Gimenez-Grau:2019hez,Gimenez-Grau:2022czc,Dey:2024ilw}. A tantalizing possibility is that requiring the $O(3)$ WF CFT to support stable Bose-Kondo impurities shrinks the $O(3)$ WF \emph{bulk} bootstrap island \cite{Kos:2015mba,Kos:2016ysd,Chester:2020iyt}. A promising setup would augment the bulk $O(3)$ WF bootstrap by including the $S=1/2$ Bose-Kondo impurity endpoints, since these endpoint primaries have the smallest scaling dimension among the Bose-Kondo impurities (c.f. Fig.~\ref{fig:defcr}) and sit above the local $O(3)$ vector primary in the bootstrap hierarchy as the $S=1/2$ fusion rules permit the endpoints to fuse into the vector. 

\emph{Acknowledgments.} This work was performed using the FuzzifiED \cite{Fuzzified} Julia package for fuzzy sphere numerics. A.S. is grateful to Roger Melko for hosting him at Perimeter Institute during the time that this work was conducted. The authors are also grateful to Roger Melko for providing computing resources that facilitated this work, and to Davide Gaiotto, Zohar Komargodski, Brandon Rayhaun, and Cenke Xu for helpful discussions. Research at Perimeter Institute is supported in part by the Government of Canada through the Department of Innovation, Science and Industry Canada and by the Province of Ontario through the Ministry of Colleges and Universities. Y.C.H. thanks IHES for its hospitality during the completion of this work.

\let\oldaddcontentsline\addcontentsline
\renewcommand{\addcontentsline}[3]{}
\bibliography{bibliography,bib_fuzzy}
\let\addcontentsline\oldaddcontentsline

\onecolumngrid
\changefontsize[16.5]{11}

\appendix

\tableofcontents

\section{Numerical Details}
\label{app:num_dets}

For the realization of the bulk $O(3)$ WF CFT, refer to Refs.~\cite{Dey2025,Guo2025}. We express $U(r), V(r)$ in the bulk model in terms of Haldane pseudopotentials $U_l,V_l$ and take $U_0 = 6.5$, $U_1 = 1.0$, $V_0 = 18.2$, $V_1 = 2.8$. The bulk critical point is found at $h=29.984 + 13.18 R^{-2.16}$. We choose the defect coupling $J_{\mathrm{imp}}=320, 250, 190$ for $S=1/2, 1, 3/2$ respectively.

\begin{figure}[htbp]
    \includegraphics[width=0.43\textwidth]{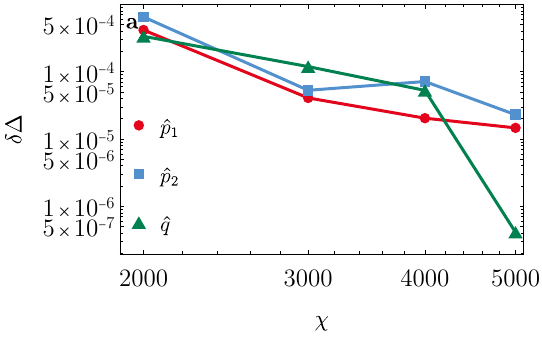} 
    \includegraphics[width=0.43\textwidth]{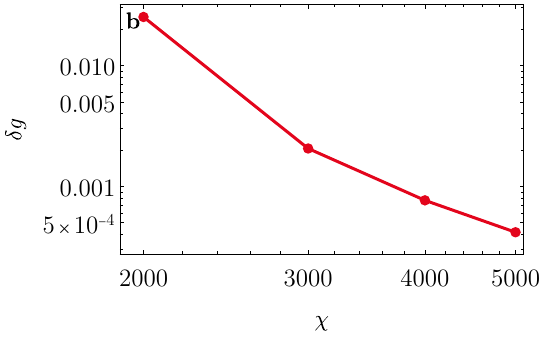} 
    \caption{DMRG Convergence, $S=1/2$ Bose-Kondo Defect, $N_m=15$. We take measurements of the optimized states at $\chi=1000,2000,3000,4000,5000$. Here we calibrate $\Delta_{\hat{O}} (\chi_i)= 2\frac{E_{\hat{O}}(\chi_i) - E_{\hat{\mathbbold{1}}}(\chi_{i})}{E_{\hat{D}}(\chi_i) - E_{\hat{\mathbbold{1}}}(\chi_i)}$, and define $\delta \Delta_{\hat{O}}(\chi_i) = \Delta_{\hat{O}}(\chi_i) - \Delta_{\hat{O}}(\chi_{i-1})$, $\delta g(\chi_i) = g(\chi_i) - g(\chi_{i-1})$.}
    \label{fig:dmrg_conv}
\end{figure}

For the DMRG calculations we utilize maximum bond dimension $\chi=5000$. From the difference of the converged energy values with varying bond dimension we estimate that the scaling dimensions have convergence error $\lesssim 10^{-4}$ on the largest system size $N_m=15$. Likewise the extracted $g$-function on $N_m=15$ has convergence error $\lesssim 10^{-3}$. The convergence with bond dimension is shown in Fig.~\ref{fig:dmrg_conv}. These errors are negligible compared to the finite-size effect. 

\section{Calculation of the $g$-Function and Defect-Creation Operators}
\label{app:g_fac}

In Ref.~\cite{Zhou2024Jan}, it is explained that the $g$-function can be extracted from the overlaps of the ground states of various defect configurations,
\begin{equation}
    g_a = \frac{|\langle a0 | 00 \rangle|^2}{|\langle a0 | aa \rangle|^2}
\end{equation}
where $\ket{{ab}}$ is the lowest state where defect $a$ lives at the upper half-infinite line and $b$ lives at the lower half-infinite line, or north and south pole after radial quantization. More concretely, $\ket{00}$ is the bulk ground state, $\ket{aa}$ is the defect ground state, and $\ket{a0}$ is the lowest defect creation operator. This applies to the case where the the Hilbert space in the presence of the defect is the same as the bulk. However, in our case, $\ket{00}, \ket{a0}, \ket{aa}$ live in different Hilbert spaces and their overlaps cannot be directly taken. 

To take the overlap between $\ket{00}\in\mathcal{H}_{00}$ and $\ket{a0}\in\mathcal{H}_{a0}$, as $\mathcal{H}_{a0}=\mathcal{H}_{00}\otimes\mathbb{C}^{2S+1}$, we direct product the bulk ground state with a spin-$S$ state at the north pole, and take
\begin{equation*}
    \langle a0,m|\left(|00\rangle\otimes|m\rangle_N\right)
\end{equation*}
as the lowest defect-creation operator has spin-$S$, the states are marked by the magnetic quantum number $m$. Similarly, to take the overlap between $\ket{aa}$ and $\ket{a0}$, we direct product $\ket{a0}$ with a spin-$S$ state at the south pole. In this way, the $g$-function is given by
\begin{equation}
    g=\frac{\sum_{m}\left|\langle a0,m|\left(|00\rangle\otimes|m\rangle_N\right)\right|^2}{\sum_{m}\left|\left(\langle a0,-m|\otimes\langle m|_S\right)|aa\rangle\right|^2}
\end{equation}
In the decoupled limit in the UV, this expression gives $g_\text{UV}=2S+1$.

\begin{figure}[htbp]
    \centering
    \includegraphics[width=0.5\linewidth]{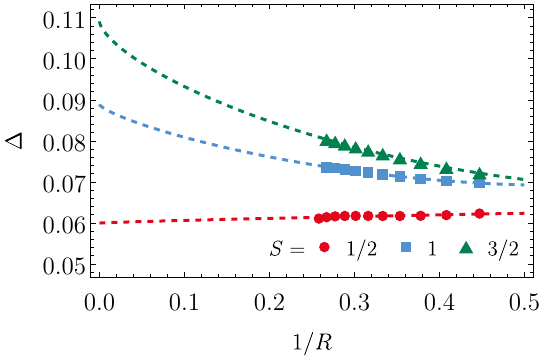}
    \caption{The finite-size scaling of the leading defect-creation operator in the Bose-Kondo defect with $S=1/2,1$ and $3/2$.}
    \label{fig:defcr}
\end{figure}

Within this framework, we have also extracted the scaling dimension of the lowest defect-creation operator through state-operator correspondence~\cite{Zhou2024Jan}
\begin{equation}
    \Delta_{a0}=\frac{R}{v}\left(E_{a0}-\tfrac{1}{2}(E_{00}+E_{aa})\right)
\end{equation}
We find the lowest defect-creation operator carries $SO(3)_\text{int}$ spin-$S$ for $S=1/2,1,3/2$ and extract its scaling dimension through finite-size scaling (Fig.~\ref{fig:defcr})
\begin{alignat}{4}
    &\Delta_{a0,S=1/2}&=0.060+0&.0043&R^{-0.82}-0&.00024&R^{-2.80}\nonumber\\
    &\Delta_{a0,S=1}&=0.089-0&.044&R^{-0.74}+0&.036&R^{-2.38}\nonumber\\
    &\Delta_{a0,S=3/2}&=0.109-0&.072&R^{-0.65}+0&.036&R^{-2.30}
\end{alignat}

\section{$O(2)$ WF Spin Impurities --- The Halon Defect}
\label{app:Halon}

Similarly to the $O(3)$ Bose-Kondo impurities, we can couple a spin-$S$ impurity to bulk $O(2)$ criticality in a manner that preserves the $O(2)_{\mathrm{int}}$ symmetry of the bulk. For $S=1/2$, anomaly arguments~\cite{Defect_anomalies} imply that for $\gamma=0$ the Hamiltonian flows to a nontrivial conformal defect. For $S\geq1$ and higher cases, further relevant couplings like $\gamma_2 (S_{N,z}^2 + S_{S,z}^2)$ are allowed, so the Halon defects become unstable and require fine tuning. For simplicity we only consider the $S=1/2$ case.

To investigate these defects we utilize the fuzzy sphere model of \cite{Guo2025} to realize the $O(2)$ WF fixed point in the bulk. For the $S=1/2$ impurity, the most general Hamiltonian that preserves $O(2)_{\mathrm{int}}$ and time-reversal symmetry is
\begin{equation}
\label{eqn:halon_ham}
    H = H_{\mathrm{bulk}} + J_{\mathrm{imp}} (V(\rN) S_\rN^- + \bar{V}(\rN) S_\rN^+ - V(\rS) S_\rS^- - \bar{V}(\rS) S_\rS^+)
\end{equation} 
where $V=\frac{1}{2}(V_x+iV_y)$ is the $O(2)$ order parameter in the bulk, and $\bar{V}=\frac{1}{2}(V_x-iV_y)$~\cite{Guo2025}. We take the defect coupling as $J_{\mathrm{imp}}=0.5$. 

We indeed observe integer-spaced conformal multiplets, indicating a nontrivial conformal defect CFT, we observe a single relevant pseudoscalar $\hat{\chi}$ (odd under the improper $\mathbb{Z}_2$ of $O(2)_\text{int}$) whose scaling dimension observed at finite size $N_m=8$ is $\Delta_{\hat{\chi},N_m=8}\approx 0.4$. The rest of the low-energy spectrum is similar to the $S=1/2$ Bose-Kondo defect. In particular, the lowest nontrivial primary $\hat{p}_1$ has $l^z=0$ and $SO(2)_\text{int}$ charge $q=1$, and there is a primary $\hat{p}_2$ with $l^z=0, n=2$ near marginality. See Table~\ref{tab:o2_1/2_data} for the full spectrum. At finite size $N_m=8$, the $g$-function is $g_{N_m=8}\approx1.7$. We leave a more rigorous numerical study of the Halon impurities to future work. 

\section{Extraction of Scaling Dimensions and Finite Size Scaling}
\label{app:dim_fss}

We calibrate the scaling dimensions by
\begin{align}
    \Delta&=\frac{\Delta E}{E_1},&E_1&=\frac{1}{4}\left[(\Delta E_{\partial p_1}-\Delta E_{p_1})+(\Delta E_{\partial p_2}-\Delta E_{p_2})+\Delta E_D\right]
\end{align}
We perform finite-size scaling of the scaling dimensions by choosing two irrelevant scalar primaries $\hat{n}_{1/2}$ and performing a fitting for $\Delta_{n_1}(R)$, $\Delta_{n_2}(R)$, and $\Delta_{O}(R)$ with $O$ labeling other low-lying primaries according to the ansatz 
\begin{equation}
    \left.\begin{aligned}
        \Delta_{n_1}(R)&=\Delta_{n_1}+c_{n_1,n_1}R^{\Delta_{n_1}-1}+c_{n_1,n_2}R^{\Delta_{n_2}-1}\\
        \Delta_{n_2}(R)&=\Delta_{n_2}+c_{n_2,n_1}R^{\Delta_{n_1}-1}+c_{n_2,n_2}R^{\Delta_{n_2}-1}\\
        \Delta_{O}(R)&=\Delta_{O}+c_{O,n_1}R^{\Delta_{n_1}-1}+c_{O,n_2}R^{\Delta_{n_2}-1}\\
    \end{aligned}\right\}
\end{equation}
The result yields
\begin{align}
    \Delta_{p_1}&=0.21(4),&\Delta_{p_2}&=0.72(4),&\Delta_{n_1}&=1.82(4)&\Delta_{n_2}&=3.80(15).
\end{align}
We take the region with $\chi^2/\chi^2_{\min}$ as our error bar. 

\begin{figure}[htbp]
    \centering
    \includegraphics[width=0.33\linewidth]{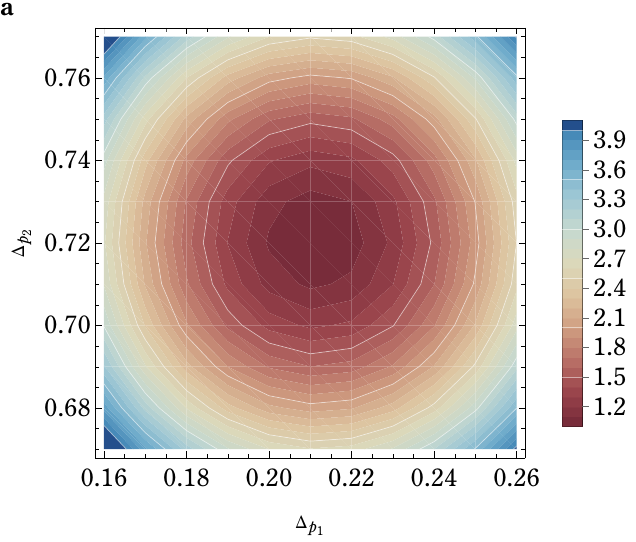}
    \includegraphics[width=0.33\linewidth]{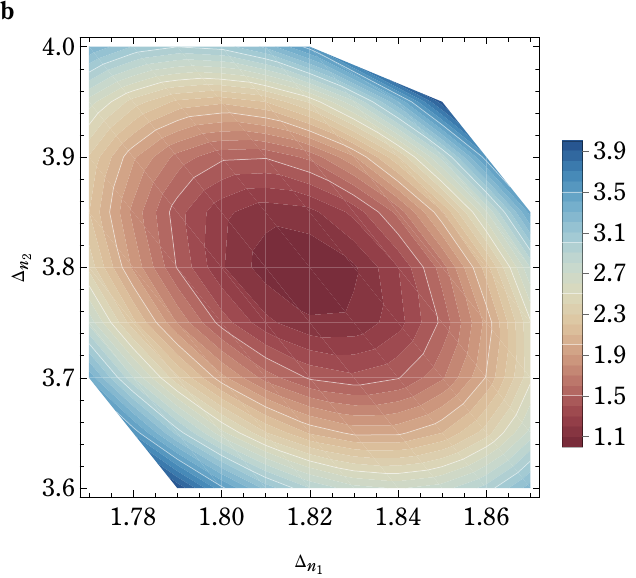}
    \caption{The fitting goodness $\chi^2/\chi^2_{\min}$ as a function of (a) $\Delta_{p_1}$ and $\Delta_{p_2}$, (b) $\Delta_{n_1}$ and $\Delta_{n_2}$. }
    \label{fig:fit}
\end{figure}

For the $g$-function, we try three different fittings 
\begin{align}
    g(R)&=g+C_1R^{-1}\nonumber\\
    g(R)&=g+C_1R^{-1}+C_2R^{-2}\nonumber\\
    g(R)&=g+C_1R^{-1}+C_2R^{-2}+C_3R^{-3}
\end{align}
and take their standard deviation as the error bar. The result yields 
\begin{align}
    g_{S=1/2}&=1.47(5),&g_{S=1}&=2.01(13),&g_{S=3/2}&=2.59(23)
\end{align}

\section{Full Spectrum Data}
\label{app:data}

\begin{figure}
    \includegraphics[width=0.32\linewidth]{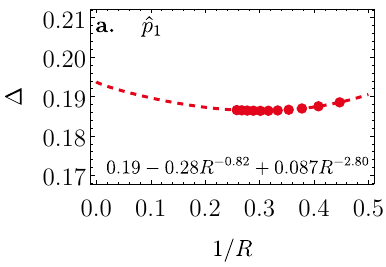} 
    \includegraphics[width=0.32\linewidth]{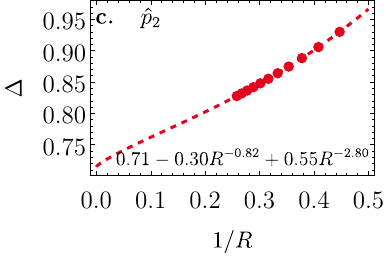} 
    \includegraphics[width=0.32\linewidth]{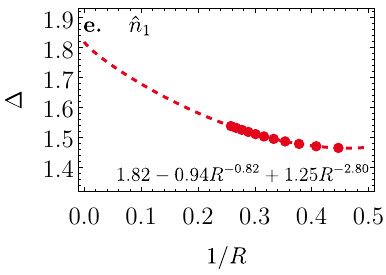} 
    \includegraphics[width=0.32\linewidth]{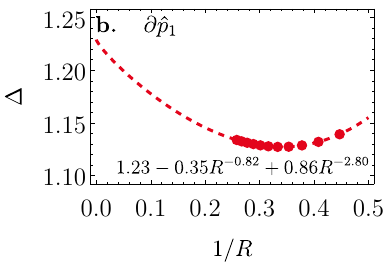} 
    \includegraphics[width=0.32\linewidth]{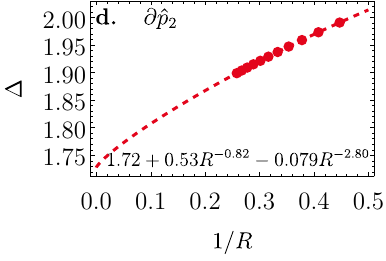} 
    \includegraphics[width=0.32\linewidth]{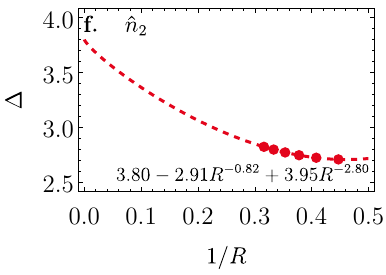} 
    \caption{Finite-Size scaling of various scaling dimensions of the $S=1/2$ Bose-Kondo defect. For $\hat{n}_2$, only ED data are available.}
    \label{fig:finite_size_scaling}
\end{figure}

\begin{figure*}[t!]
    \includegraphics[width=0.32\linewidth]{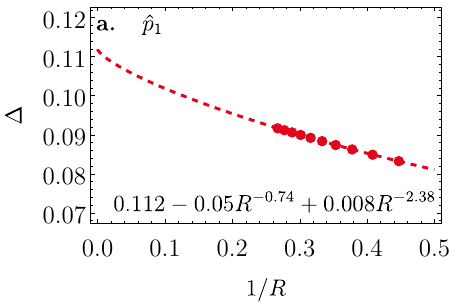} 
    \includegraphics[width=0.32\linewidth]{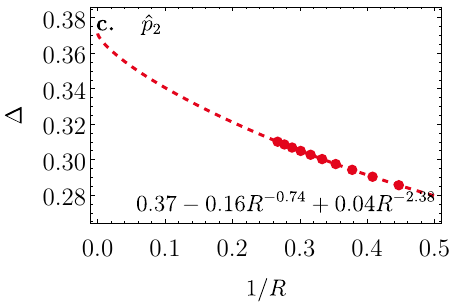} 
    \includegraphics[width=0.32\linewidth]{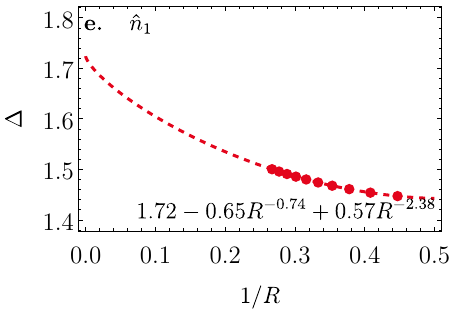} 
    \includegraphics[width=0.32\linewidth]{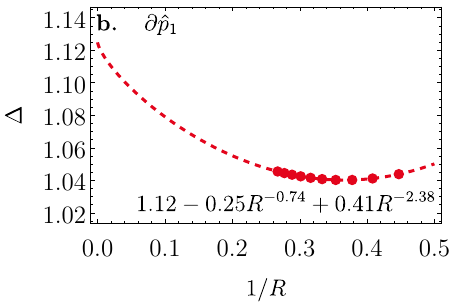} 
    \includegraphics[width=0.32\linewidth]{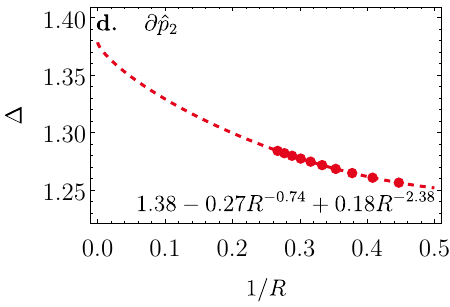} 
    \includegraphics[width=0.32\linewidth]{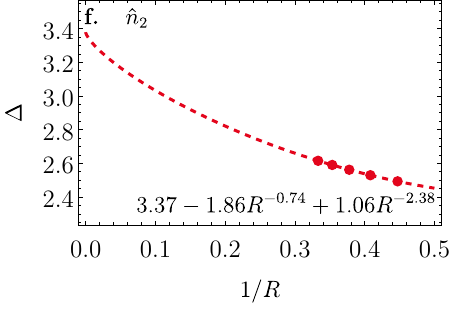} 
    \caption{Finite-Size scaling of various scaling dimensions of the $S=1$ Bose-Kondo defect. For $\hat{n}_2$, only ED data are available.}
    \label{fig:finite_size_scaling_1}
\end{figure*}

\begin{figure*}[t!]
    \includegraphics[width=0.32\linewidth]{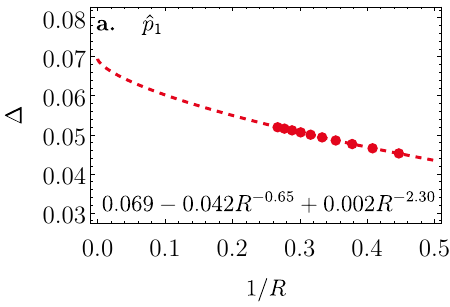} 
    \includegraphics[width=0.32\linewidth]{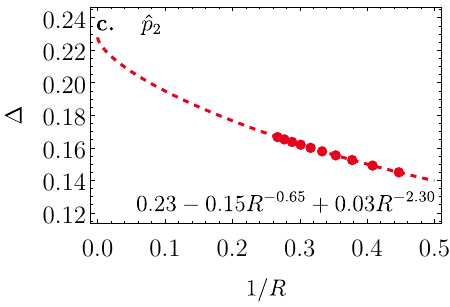} 
    \includegraphics[width=0.32\linewidth]{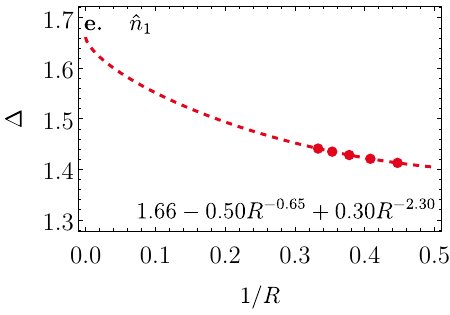} 
    \includegraphics[width=0.32\linewidth]{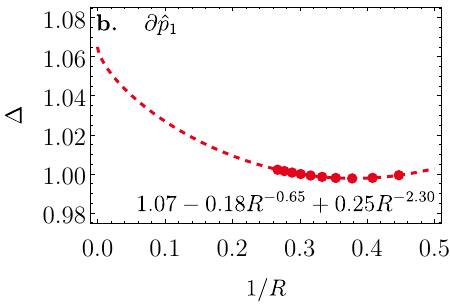} 
    \includegraphics[width=0.32\linewidth]{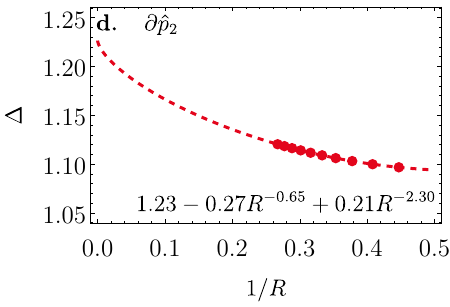} 
    \includegraphics[width=0.32\linewidth]{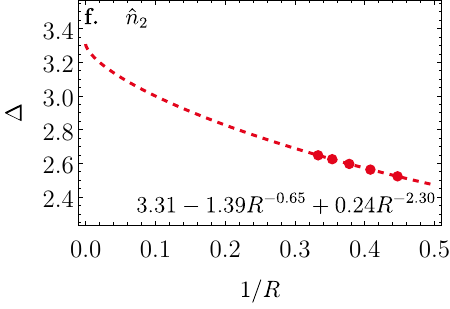} 
    \caption{Finite-Size scaling of various scaling dimensions of the $S=3/2$ Bose-Kondo defect. For $\hat{n}_2$, only ED data are available.}
    \label{fig:finite_size_scaling_3/2}
\end{figure*}

In Tables~\ref{tab:o3_1/2_ed_spectrum}--\ref{tab:o2_1/2_data} we show our ED and DMRG results on the largest accessible system sizes for the low-lying spectra of the $S=1/2,1,3/2$ Bose-Kondo and $S=1/2$ Halon defects. In each case clear conformal multiplet structure can be seen with approximate integer spacing of scaling dimensions. We emphasize that this data is subject to finite-size corrections. In Figures \ref{fig:finite_size_scaling}--\ref{fig:finite_size_scaling_3/2} we show the fits of the spectra of the Bose-Kondo defects with the prediction from conformal perturbation theory.


\begin{table}[htbp]
    \centering
    \renewcommand{\tabcolsep}{10pt}
    \begin{tabular}{c|cc|ll}
        \hline\hline
        $\hat{O}$&$l^z$&$s$&$\Delta_{\hat{O}}$&$\Delta_{\partial \hat{O}}-\Delta_{\hat{O}}$\\
        \hline
        $\hat{p}_1$ & $0$ & $1$ & $0.21$ & $1.04$ \\
        $\hat{p}_2$ & $0$ & $2$ & $0.72$ & $1.01$ \\
        $\hat{q}$ & $0$ & $1$ & $1.03$ & $1.06^\dagger$ \\
        & $0$ & $3$ & $1.40$ & $1.13^\ast$ \\
        & $1$ & $1$ & $1.58$ & $1.00^\dagger$ \\
        & $1$ & $1$ & $1.68$ & $1.00^\dagger$ \\
        $\hat{S}/\hat{n}_1$ & $0$ & $0$ & $1.82$ & $0.98^\ast$ \\
        $\hat{D}$ & $1$ & $0$ & $1.95$ & $1.07^\ast$\\
        $\hat{n}_2$ & $0$ & $0$ & $3.80$ & / \\
        \hline\hline
    \end{tabular}
    \caption{Low-Lying Primaries, $S=1/2$. We report the values after extrapolated from finite-size scaling for most scaling dimensions, except that the values marked with ``$\ast$'' are finite-size values at the largest ED-accessible size $N_m=10$, and the values marked with ``$\dagger$'' are finite-size values at the largest DMRG-accessible size $N_m=15$. }
    \label{tab:1/2_prims}
\end{table}

\begin{table}[htbp]
    \centering
    \renewcommand{\tabcolsep}{10pt}
    \begin{tabular}{c|cc|ll}
        \hline\hline
        $\hat{O}$&$l^z$&$s$&$\Delta_{\hat{O}}$&$\Delta_{\partial \hat{O}}-\Delta_{\hat{O}}$\\
        \hline
        $\hat{p}_1$&0&1&$0.12$&$1.01$\\
        $\hat{p}_2$&0&2&$0.37$&$1.00$\\
        $\hat{p}_3$&0&3&$0.93$&$1.10^\ast$\\
        $\hat{q}$&0&1&$1.06$&$1.18^\ast$\\
        &0&2&$1.13$&$1.03^\ast$\\
        $\hat{S}/\hat{n}_1$&0&0&$1.73$&$1.17^\ast$\\
        &0&1&$1.86$&$1.15^\ast$\\
        $\hat{D}$&1&0&$1.98$&$1.07^\ast$\\
        $\hat{n}_2$&0&0&$3.38$&/\\
        \hline\hline
    \end{tabular}
    \caption{Low-Lying Primaries, $S=1$. Values marked with ``$\ast$'' are finite-size values at the largest ED-accessible size $N_m=9$. }
    \label{tab:1_prims}
\end{table}

\begin{table}[htbp]
    \centering
    \renewcommand{\tabcolsep}{10pt}
    \begin{tabular}{c|cc|ll}
        \hline\hline
        $\hat{O}$&$l^z$&$s$&$\Delta_{\hat{O}}$&$\Delta_{\partial \hat{O}}-\Delta_{\hat{O}}$\\
        \hline
        $\hat{p}_1$&0&1&$0.07$&$1.00$\\
        $\hat{p}_2$&0&2&$0.23$&$1.00$\\
        $\hat{p}_3$&0&3&$0.50$&$0.96^\ast$\\
        $\hat{q}$&0&1&$1.04$&$1.21^\ast$\\
        $\hat{p}_4$&0&4&$1.08$&$1.08^\ast$\\
        &0&2&$1.12$&$1.05^\ast$\\
        $\hat{S}/\hat{n}_1$&0&0&$1.65$&$1.22^\ast$\\
        $\hat{D}$&1&0&$2.01$&$0.99^\ast$\\
        $n_2$&0&0&$3.30$&/\\
        \hline\hline
    \end{tabular}
    \caption{Low-Lying Primaries, $S=3/2$. Values marked with ``$\ast$'' are finite-size values at the largest ED-accessible size $N_m=9$. }
    \label{tab:3/2_prims}
\end{table}

\begin{table}
    \renewcommand{\tabcolsep}{10pt}
    \begin{tabular}{c c c c c | c c c c c | c c c c c}
    \hline \hline
    Op. & $\Delta$ & $l^z$ & $s$ & $P$ &
    Op. & $\Delta$ & $l^z$ & $s$ & $P$ &
    Op. & $\Delta$ & $l^z$ & $s$ & $P$ \\
    \hline
    $\hat{\mathbbold{1}}$ & 0 & 0 & 0 & 1   &  & 1.846 & 1 & 1 &  &  & 2.696 & 1 & 0 & \\
    \cline{1-5}\cline{11-15}
    $\hat{p}_1$ & 0.186 & 0 & 1 & -1  &  & 2.819 & 1 & 1 &  & $\hat{n}_2$ & 2.823 & 0 & 0 & 1 \\
    \cline{6-10}\cline{11-15}
    $\p \hat{p}_1$ & 1.128 & 0 & 1 & 1   & $\hat{D}$ & 1.985 & 1 & 0 &  &  & 2.839 & 2 & 1 & \\
    \cline{11-15}
    & 2.059 & 0 & 1 & -1  &  & 3.051 & 1 & 0 &  &  & 2.880 & 0 & 1 & 1 \\
    \cline{6-10}\cline{11-15}
    & 2.992 & 0 & 1 & 1   &  & 1.988 & 0 & 2 & 1   &  & 2.887 & 0 & 1 & -1 \\
    \cline{1-5}\cline{11-15}
    $\hat{p}_2$ & 0.855 & 0 & 2 & 1   &  & 3.100 & 0 & 2 & -1  &  & 2.927 & 0 & 4 & 1 \\
    \cline{6-10}\cline{11-15}
    $\p \hat{p}_2$ & 1.929 & 0 & 2 & -1  &  & 2.005 & 1 & 2 &   &  & 2.930 & 2 & 1 & \\
    \cline{11-15}
    & 2.946 & 0 & 2 & 1   &  & 3.018 & 1 & 2 &   &  & 2.955 & 1 & 3 & \\
    \cline{1-5}\cline{6-10}\cline{11-15}
    $\hat{q}$ & 1.027 & 0 & 1 & -1  &  & 2.139 & 0 & 0 & 1   &  & 2.964 & 1 & 2 & \\
    \cline{11-15}
    & 2.134 & 0 & 1 & 1   &  & 3.221 & 0 & 0 & -1  &  & 2.978 & 1 & 1 & \\
    \cline{6-10}\cline{11-15}
    & 3.205 & 0 & 1 & -1  &  & 2.261 & 0 & 1 & -1  &  & 3.073 & 2 & 0 & \\
    \cline{1-5}\cline{11-15}
    $\hat{S}/\hat{n}_1$ & 1.503 & 0 & 0 & 1   &  & 3.293 & 0 & 1 & 1   &  & 3.105 & 1 & 0 & \\
    \cline{6-10}\cline{11-15}
    & 2.483 & 0 & 0 & -1  &  & 2.343 & 0 & 2 & -1  &  & 3.120 & 0 & 3 & -1 \\
    \cline{1-5}\cline{6-10}\cline{11-15}
    & 1.758 & 1 & 1 &   &  & 2.352 & 0 & 1 & -1  &  & 3.132 & 2 & 2 & \\
    \cline{6-10}\cline{11-15}
    & 2.762 & 1 & 1 &   &  & 2.462 & 1 & 2 &   &  & 3.158 & 1 & 1 & \\
    \cline{1-5}\cline{6-10}\cline{11-15}
    & 1.781 & 0 & 3 & -1  &  & 2.619 & 1 & 1 &   &  & 3.194 & 1 & 2 & \\
    \cline{6-10}\cline{11-15}
    & 2.911 & 0 & 3 & 1   &  & 2.624 & 0 & 2 & 1   &  & 3.232 & 1 & 1 & \\
    \cline{1-5}\cline{6-10}\cline{11-15}
    \hline \hline
    \end{tabular}
    \caption{Spin-$1/2$ Bose-Kondo Spectrum, ED $N_m=10$. 51 lowest energy states are shown. Conformal multiplets are separated by horizontal lines. We label certain operators of interest.}
    \label{tab:o3_1/2_ed_spectrum}
\end{table}

\begin{table}
    \renewcommand{\tabcolsep}{10pt}
    \begin{tabular}{c c c c c | c c c c c | c c c c c}
    \hline\hline
    Op. & $\Delta$ & $l^z$ & $s$ & $P$ &
    Op. & $\Delta$ & $l^z$ & $s$ & $P$ &
    Op. & $\Delta$ & $l^z$ & $s$ & $P$ \\
    \hline
    $\hat{\mathbbold{1}}$ & 0 & 0 & 0 & 1   &  & 1.729 & 1 & 1 &   &  & 2.493 & 0 & 1 & 1 \\
    \cline{1-5}\cline{11-15}
    $\hat{p}_1$ & 0.088 & 0 & 1 & -1  &  & 2.715 & 1 & 1 &   &  & 2.525 & 0 & 1 & -1 \\
    \cline{6-10}\cline{11-15}
    $\p \hat{p}_1$ & 1.041 & 0 & 1 & 1   &  & 1.856 & 1 & 1 &   &  & 2.535 & 0 & 2 & 1 \\
    \cline{11-15}
    & 2.153 & 0 & 1 & -1  &  & 2.847 & 1 & 1 &   &  & 2.579 & 0 & 3 & 1 \\
    \cline{1-5}\cline{6-10}\cline{11-15}
    $\hat{p}_2$ & 0.300 & 0 & 2 & 1   &  & 1.952 & 1 & 2 &   & $\hat{n}_2$ & 2.614 & 0 & 0 & 1 \\
    \cline{11-15}
    $\p \hat{p}_2$ & 1.271 & 0 & 2 & -1  &  & 2.958 & 1 & 2 &   &  & 2.656 & 1 & 0 & \\
    \cline{6-10}\cline{11-15}
    & 2.422 & 0 & 2 & 1   &  & 1.983 & 0 & 4 & 1   &  & 2.673 & 1 & 3 & \\
    \cline{1-5}\cline{6-10}\cline{11-15}
    $\hat{q}$ & 1.006 & 0 & 1 & -1  &  & 2.021 & 1 & 2 &   &  & 2.766 & 1 & 1 & \\
    \cline{6-10}\cline{11-15}
    & 2.186 & 0 & 1 & 1   & $\hat{D}$ & 2.076 & 1 & 0 &   &  & 2.777 & 1 & 2 & \\
    \cline{1-5}\cline{6-10}\cline{11-15}
    $\hat{p}_3$ & 1.006 & 0 & 3 & -1  &  & 2.115 & 1 & 1 &   &  & 2.803 & 2 & 1 & \\
    \cline{6-10}\cline{11-15}
    & 2.103 & 0 & 3 & 1   &  & 2.124 & 0 & 2 & 1   &  & 2.825 & 1 & 2 & \\
    \cline{1-5}\cline{6-10}\cline{11-15}
    & 1.125 & 0 & 2 & 1   &  & 2.132 & 0 & 3 & -1  &  & 2.847 & 0 & 3 & -1 \\
    \cline{6-10}\cline{11-15}
    & 2.156 & 0 & 2 & -1  &  & 2.141 & 1 & 3 &   &  & 2.869 & 0 & 1 & -1 \\
    \cline{1-5}\cline{6-10}\cline{11-15}
    $\hat{S} / \hat{n}_1$ & 1.474 & 0 & 0 & 1   &   & 2.233 & 0 & 0 & 1   &  & 2.871 & 0 & 0 & 1 \\
    \cline{6-10}\cline{11-15}
    & 2.648 & 0 & 0 & -1  &  & 2.265 & 0 & 2 & 1   &  & 2.943 & 2 & 1 & \\
    \cline{1-5}\cline{6-10}\cline{11-15}
    & 1.512 & 0 & 1 & -1  &  & 2.337 & 0 & 1 & -1  &  & 2.999 & 0 & 2 & -1 \\
    \cline{6-10}\cline{11-15}
    & 2.657 & 0 & 1 & 1   &  & 2.368 & 0 & 2 & -1  &  & 3.011 & 1 & 1 & \\
    \cline{1-5}\cline{6-10}\cline{11-15}
    \hline\hline
    \end{tabular}
    \caption{Spin-$1$ Bose-Kondo Spectrum, ED $N_m=9$. 51 lowest energy states are shown. }
\end{table}

\begin{table}
    \renewcommand{\tabcolsep}{10pt}
    \begin{tabular}{c c c c c | c c c c c | c c c c c}
    \hline\hline
    Op. & $\Delta$ & $l^z$ & $s$ & $P$ &
    Op. & $\Delta$ & $l^z$ & $s$ & $P$ &
    Op. & $\Delta$ & $l^z$ & $s$ & $P$ \\
    \hline
    $\hat{\mathbbold{1}}$ & 0 & 0 & 0 & 1    & $\hat{S}/\hat{n}_1$ & 1.441 & 0 & 0 & 1    &  & 2.167 & 1 & 2 & \\
    \cline{1-5}\cline{6-10}\cline{11-15}
    $\hat{p}_1$ & 0.049 & 0 & 1 & -1   &  & 1.454 & 0 & 1 & -1   &  & 2.169 & 0 & 3 & -1 \\
    \cline{11-15}
    $\p \hat{p}_1$ & 0.998 & 0 & 1 & 1    &  & 2.431 & 0 & 1 & 1    &  & 2.175 & 1 & 4 & \\
    \cline{6-10}\cline{11-15}
    & 2.167 & 0 & 1 & -1   &  & 1.476 & 0 & 2 & 1    &  & 2.249 & 0 & 0 & 1 \\
    \cline{1-5}\cline{11-15}
    $\hat{p}_2$ & 0.158 & 0 & 2 & 1    &  & 2.486 & 0 & 2 & -1   &  & 2.276 & 0 & 2 & -1 \\
    \cline{6-10}\cline{11-15}
    $\p \hat{p}_2$ & 1.109 & 0 & 2 & -1   &  & 1.734 & 1 & 1 &    &  & 2.314 & 0 & 1 & -1 \\
    \cline{11-15}
    & 2.201 & 0 & 2 & 1    &  & 2.661 & 1 & 1 &    &  & 2.391 & 0 & 1 & -1 \\
    \cline{1-5}\cline{6-10}\cline{11-15}
    $\hat{p}_3$ & 0.358 & 0 & 3 & -1   &  & 1.843 & 1 & 1 &    &  & 2.411 & 0 & 3 & -1 \\
    \cline{6-10}\cline{11-15}
    & 1.319 & 0 & 3 & 1    &  & 1.844 & 1 & 2 &    &  & 2.431 & 0 & 2 & 1 \\
    \cline{6-10}\cline{11-15}
    & 2.317 & 0 & 3 & -1   &  & 1.915 & 1 & 2 &    &  & 2.444 & 0 & 3 & 1 \\
    \cline{1-5}\cline{6-10}\cline{11-15}
    $\hat{q}$ & 0.980 & 0 & 1 & -1   &  & 2.016 & 1 & 3 &    &  & 2.522 & 0 & 2 & 1 \\
    \cline{6-10}\cline{11-15}
    & 2.194 & 0 & 1 & 1    &  & 2.045 & 0 & 5 & -1   &  & 2.622 & 0 & 1 & -1 \\
    \cline{1-5}\cline{6-10}\cline{11-15}
    & 1.047 & 0 & 2 & 1    & $\hat{D}$ & 2.100 & 1 & 0 &    &  & 2.627 & 1 & 0 & \\
    \cline{6-10}\cline{11-15}
    & 2.095 & 0 & 2 & -1   &  & 2.103 & 0 & 2 & 1    &  & 2.631 & 0 & 4 & -1 \\
    \cline{1-5}\cline{6-10}\cline{11-15}
    $\hat{p}_4$ & 1.065 & 0 & 4 & 1    &  & 2.116 & 1 & 3 &    &  & 2.641 & 0 & 3 & -1 \\
    \cline{6-10}\cline{11-15}
    & 2.148 & 0 & 4 & -1   &  & 2.122 & 1 & 1 &    & $\hat{n}_2$ & 2.647 & 0 & 0 & 1 \\
    \cline{1-5}\cline{6-10}
    & 1.151 & 0 & 3 & -1   &  & 2.157 & 0 & 4 & 1    &  &       &   &   &    \\
    & 2.207 & 0 & 3 & 1    &  &       &   &   &      &  &       &   &   &    \\
    \hline\hline
    \end{tabular}
    \caption{Spin-3/2 Bose-Kondo Spectrum, ED $N_m=9$. 51 lowest energy states are shown. }
\end{table}

\begin{table}[h!]
    \renewcommand{\tabcolsep}{10pt}
    \begin{tabular}{c c c c c | c c c c c | c c c c c}
    \hline\hline
    Op. & $\Delta$ & $l^z$ & $s$ & $P$ &
    Op. & $\Delta$ & $l^z$ & $s$ & $P$ &
    Op. & $\Delta$ & $l^z$ & $s$ & $P$\\
    \hline
    $\hat{\mathbbold{1}}$ & 0 & 0 & 0  & 1 &  & 1.732 & 1 & 1  &  &  & 2.84 & 2 & 1 & \\
    \cline{1-5}\cline{11-15}
    $\hat{p}_1$ & 0.187 & 0 & 1  & -1 &  & 2.732 & 1 & 1  &  &  & 2.856 & 1 & 3 & \\
    \cline{6-10}\cline{11-15}
    $\p \hat{p}_1$ & 1.134 & 0 & 1  & 1 &  & 1.809 & 1 & 1  &  &  & 2.916 & 2 & 1 & \\
    \cline{1-5}\cline{11-15}
    $\hat{p}_2$ & 0.827 & 0 & 2  & 1 &  & 2.81 & 1 & 1  &  &  & 3.08 & 2 & 0 & \\
    \cline{6-10}\cline{11-15}
    $\p \hat{p}_2$ & 1.899 & 0 & 2  & -1 &  & 1.948 & 0 & 2  & 1 &  & 3.091 & 2 & 2 & \\
    \cline{1-5}\cline{6-10}\cline{11-15}
    $\hat{q}$ & 1.018 & 0 & 1  & -1 &  & 1.958 & 1 & 2  &  &  & 3.367 & 2 & 2 & \\
    \cline{6-10}\cline{11-15}
    & 2.077 & 0 & 1  & 1 &$\hat{D}$ & 1.981 & 1 & 0  &  &  & 3.42 & 2 & 0 & \\
    \cline{1-5}\cline{6-10}\cline{11-15}
    $\hat{S}/\hat{n}_1$ & 1.538 & 0 & 0  & 1 &  & 2.446 & 1 & 2  &  &  & 3.489 & 2 & 1 & \\
    \cline{1-5}\cline{6-10}\cline{11-15}
    & 1.709 & 0 & 3  & -1 &  & 2.63 & 1 & 1  &  &  & 3.498 & 2 & 2 & \\
    \cline{6-10}\cline{11-15}
    &       &   &  &  &  & 2.699 & 1 & 0  &  &  & 3.53 & 2 & 1 & \\
    \cline{11-15}
    &       &   &  &  &  &       &   &  &  &  & 3.565 & 2 & 2  & \\
    \cline{11-15}
    \hline\hline
\end{tabular}

\caption{Spin-1/2 Bose-Kondo Spectrum, DMRG $N_m=15$. 10 lowest energy states in each $l^z=0,1,2$ sector are shown.}
\end{table}

\begin{table}
    \renewcommand{\tabcolsep}{10pt}
    \begin{tabular}{c c c c c | c c c c c | c c c c c}
    \hline\hline
    Op. & $\Delta$ & $l^z$ & $s$ & $P$ &
    Op. & $\Delta$ & $l^z$ & $s$ & $P$ &
    Op. & $\Delta$ & $l^z$ & $s$ & $P$ \\
    \hline
    $\hat{\mathbbold{1}}$ & 0 & 0 & 0 & 1   &  & 1.835 & 1 & 1 &   &  & 3.080 & 2 & 2 & \\
    \cline{1-5}\cline{11-15}
    $\hat{p}_1$ & 0.092 & 0 & 1 & -1  &  & 2.744 & 1 & 1 &   &  & 3.148 & 2 & 2 & \\
    \cline{6-10}\cline{11-15}
    $\p \hat{p}_1$ & 1.045 & 0 & 1 & 1   &  & 1.937 & 1 & 2 &   &  & 3.180 & 2 & 0 & \\
    \cline{1-5}\cline{6-10}\cline{11-15}
    $\hat{p}_2$ & 0.310 & 0 & 2 & 1   &  & 2.022 & 1 & 2 &   &  & 3.265 & 2 & 1 & \\
    \cline{6-10}\cline{11-15}
    $\p \hat{p}_2$ & 1.284 & 0 & 2 & -1  & $\hat{D}$ & 2.072 & 1 & 0 &   &  & 3.299 & 2 & 3 & \\
    \cline{1-5}\cline{6-10}\cline{11-15}
    $\hat{p}_3$ & 0.993 & 0 & 3 & -1  &  & 2.127 & 1 & 1 &   &  & 3.476 & 2 & 0 & \\
    \cline{1-5}\cline{6-10}\cline{11-15}
    $\hat{q}$ & 1.007 & 0 & 1 & -1  &  & 2.135 & 1 & 3 &   &  & 3.515 & 2 & 2 & \\
    \cline{1-5}\cline{6-10}\cline{11-15}
    & 1.120 & 0 & 2 & 1   &  & 2.692 & 1 & 3 &   &  & 3.550 & 2 & 1 & \\
    \cline{1-5}\cline{6-10}\cline{11-15}
    $\hat{S}/\hat{n}_1$ & 1.500 & 0 & 0 & 1   &  & 2.887 & 2 & 1 &   &  &       &   &   &   \\
    \cline{1-5}\cline{6-10}
    & 1.553 & 0 & 1 & -1  &  & 2.972 & 2 & 1 &   &  &       &   &   &   \\
    \cline{1-5}\cline{6-10}
    & 1.754 & 1 & 1 &   &  &       &   &   &     &  &       &   &   &   \\
    & 2.689 & 1 & 1 &   &  &       &   &   &     &  &       &   &   &   \\
    \cline{1-5}
    \hline\hline
    \end{tabular}
    \caption{Spin-1 Bose-Kondo Spectrum, DMRG $N_m=14$. 10 lowest energy states in each $l^z=0,1,2$ sector are shown.}
\end{table}

\begin{table}
    \renewcommand{\tabcolsep}{10pt}
    \begin{tabular}{c c c c c | c c c c c | c c c c c}
    \hline \hline
    Op. & $\Delta$ & $l^z$ & $s^z$ & $P$ &
    Op. & $\Delta$ & $l^z$ & $s^z$ & $P$ &
    Op. & $\Delta$ & $l^z$ & $s^z$ & $P$ \\
    \hline
    $\hat{\mathbbold{1}}$ & 0 & 0 & 0 & 1    &  & 1.765 & 1 & 1 &    &  & 2.903 & 2 & 1 & \\
    \cline{1-5}\cline{6-10}\cline{11-15}
    $\hat{p}_1$ & 0.052 & 0 & 1 & -1   &  & 1.835 & 1 & 1 &    &  & 2.982 & 2 & 1 & \\
    \cline{6-10}\cline{11-15}
    $\p \hat{p}_1$ & 1.002 & 0 & 1 & 1    &  & 1.875 & 1 & 2 &    &  & 3.025 & 2 & 2 & \\
    \cline{1-5}\cline{6-10}\cline{11-15}
    $\hat{p}_2$ & 0.167 & 0 & 2 & 1    &  & 1.921 & 1 & 2 &    &  & 3.061 & 2 & 2 & \\
    \cline{6-10}\cline{11-15}
    $\p \hat{p}_2$ & 1.120 & 0 & 2 & -1   &  & 2.025 & 1 & 3 &    &  & 3.177 & 2 & 3 & \\
    \cline{1-5}\cline{6-10}\cline{11-15}
    $\hat{p}_3$ & 0.376 & 0 & 3 & -1   & $\hat{D}$ & 2.096 & 1 & 0 &    &  & 3.217 & 2 & 0 & \\
    \cline{1-5}\cline{6-10}\cline{11-15}
    $\hat{q}$ & 0.983 & 0 & 1 & -1   &  & 2.125 & 1 & 1 &    &  & 3.257 & 2 & 1 & \\
    \cline{1-5}\cline{6-10}\cline{11-15}
    & 1.054 & 0 & 2 & 1    &  & 2.126 & 1 & 3 &    &  & 3.279 & 2 & 3 & \\
    \cline{1-5}\cline{6-10}\cline{11-15}
    $\hat{p}_4$ & 1.067 & 0 & 4 & 1    &  & 2.188 & 1 & 2 &    &  & 3.351 & 2 & 2 & \\
    \cline{1-5}\cline{6-10}\cline{11-15}
    & 1.156 & 0 & 3 & -1   &  & 2.194 & 1 & 4 &    &  & 3.371 & 2 & 4 & \\
    \cline{1-5}\cline{6-10}\cline{11-15}
    \hline\hline
    \end{tabular}
    \caption{Spin-3/2 Bose-Kondo Spectrum, DMRG $N_m=14$. 10 lowest energy states in each $l^z=0,1,2$ sector are shown.}
\end{table}

\begin{table}
    \renewcommand{\tabcolsep}{10pt}
    \begin{tabular}{c c c c c | c c c c c | c c c c c}
    \hline\hline
    Op. & $\Delta$ & $l^z$ & $n$ & $P$ &
    Op. & $\Delta$ & $l^z$ & $n$ & $P$ &
    Op. & $\Delta$ & $l^z$ & $n$ & $P$ \\
    \hline
    $\hat{\mathbbold{1}}$ & 0 & 0 & 0 & 1    & $\hat{D}$ & 2 & 1 & 0 &    &  & 2.543 & 0 & 0 & -1 \\
    \cline{1-5}\cline{11-15}
    & 0.129 & 0 & 1 & -1   &  & 3.010 & 1 & 0 &    &  & 2.708 & 0 & 1 & -1 \\
    \cline{6-10}\cline{11-15}
    & 1.132 & 0 & 1 & 1    &  & 2.001 & 1 & 1 &    &  & 3.020 & 0 & 0 & 1 \\
    \cline{11-15}
    & 2.105 & 0 & 1 & -1   &  & 2.980 & 1 & 1 &    &  & 3.080 & 1 & 0 &  \\
    \cline{6-10}\cline{11-15}
    & 3.005 & 0 & 1 & 1   &  & 2.109 & 1 & 2 &    &  & 3.108 & 2 & 1 &  \\
    \cline{1-5}\cline{11-15}
    $\hat{\chi}$ & 0.386 & 0 & 0 & -1   &  & 3.236 & 1 & 2 &    &  & 3.175 & 2 & 0 &  \\
    \cline{6-10}\cline{11-15}
    & 1.494 & 0 & 0 & 1    &  & 2.152 & 1 & 0 &    &  & 3.182 & 1 & 1 &  \\
    \cline{11-15}
    & 2.502 & 0 & 0 & -1   &  & 3.152 & 1 & 0 &    &  & 3.204 & 2 & 2 &  \\
    \cline{1-5}\cline{6-10}\cline{11-15}
    & 0.855 & 0 & 2 & 1    &  & 2.209 & 1 & 1 &    &  & 3.216 & 2 & 0 &  \\
    \cline{11-15}
    & 2.026 & 0 & 2 & -1   &  & 3.268 & 1 & 1 &    &  & 3.280 & 1 & 0 &  \\
    \cline{6-10}\cline{11-15}
    & 3.036 & 0 & 2 & 1    &  & 2.221 & 0 & 2 & 1    &  & 3.299 & 1 & 2 &  \\
    \cline{1-5}\cline{11-15}
    & 1.159 & 0 & 1 & -1   &  & 3.335 & 0 & 2 & -1   &  & 3.314 & 0 & 1 & 1 \\
    \cline{6-10}\cline{11-15}
    & 2.317 & 0 & 1 & 1    &  & 2.303 & 0 & 0 & 1    &  & 3.335 & 2 & 1 &  \\
    \cline{11-15}
    & 3.408 & 0 & 1 & -1   &  & 3.365 & 0 & 0 & -1   &  & 3.371 & 1 & 0 &  \\
    \cline{1-5}\cline{6-10}\cline{11-15}
    $\hat{S}$ & 1.236 & 0 & 0 & 1    &  & 2.438 & 0 & 1 & -1   &  & 3.406 & 1 & 2 &  \\
    \cline{11-15}
    & 2.137 & 0 & 0 & -1   &  & 3.455 & 0 & 1 & 1    &  & 3.422 & 0 & 0 & -1 \\
    \cline{6-10}
    & 3.284 & 0 & 0 & 1    &  & 2.438 & 0 & 2 & -1   &  &  &  &  &  \\

    &  &  &  &  &  & 3.503 & 0 & 2 & 1    &  &  &  &  &  \\
    \cline{6-10}

    \hline\hline
\end{tabular}

\caption{Spin-1/2 Halon Spectrum, ED $N_m=8$. 51 lowest energy states are shown. Here $n$ labels the $SO(2)_{\mathrm{int}}$ representation. Here we choose a simple calibration where $\Delta_{\hat{D}}=2$. The lowest scalar primary $\hat{S}$ has $\Delta >1$, showing the IR stability of the dCFT. The pseudo-scalar primary $\hat{\chi}$ has $\Delta_{\hat{\chi}}<1$ corresponding to the relevant $S_z$ perturbation described in Appendix \ref{app:Halon}.}
\label{tab:o2_1/2_data}
\end{table}

\end{document}